# Under the Conditions of Non-Agenda Ownership: Social Media Users in the 2019 Ukrainian Presidential Elections Campaign

Artem Zakharchenko[1,2 \[0000-0002-3877-8403\]], Yuliia Maksimtsova[2\[0000-0001-9829-0752\]], Valentyn Iurchenko[2\[0000-0002-6716-3715\]], Viktoriya Shevchenko[1\[0000-0003-1642-1283\]], Solomiia Fedushko[3 \[0000-0001-7548-5856\]]

[1] Institute of journalism, Taras Shevchenko National University of Kyiv, Kyiv, Ukraine
[2] Center for Content Analysis, Kyiv, Ukraine
[3] Lviv Polytechnic National University, Ukraine

artem.zakh@gmail.com, victoryshe@knu.ua,
artem.zakharchenko@ukrcontent.com,
maksimtsova.julia@gmail.com, valentin980301@gmail.com,
solomiia.s.fedushko@lpnu.ua

**Abstract.** Owing to its history and challenging circumstances, social networks community in Ukraine is a very interesting polygon for the study of communications in the constantly changing environment, especially in the political discourse. This unique environment requires three dimensions to ascertain the political position of its participant. But 2019 presidential elections made this object even more spectacular. The winner of elections comedian Volodymyr Zelenskyi reached 73% of votes without any issue ownership, with empty agenda, and this influenced the electoral content of social networks and their authors` behavior. We saw, that the issue ownership by other candidates succeeds in making their issues more salient in social networks. But the new phenomena, the non-agenda ownership, overcome any ideological influence, especially under the conditions of 'punishment mechanism' applied to 'old politicians'. Analyzing social media content and users behavior in the period between two rounds of elections, we found considerable overlaps between this campaign and the 2016 Trump campaign. We approved the widespread of 'filter bubbles', negative campaign messages, fake news and conspiracy theories. Active and powerful 'core' of Ukrainian Facebook that was responsible for the Revolution of dignity now became less significant and even turns into the huge filter bubble of active people. We also proved that manipulations and fake news in the environment of private groups may be as much powerful as in a case of 'classical' communication based around the opinion leaders.

# 1 Introduction

## 1.1 Social networks in the electoral campaigns. Mediatization and entertainatization of elections.

Scientists` assessment of the interplay between social network services (SNS) and electoral communications significantly evolved over the last years. Cooper noted in 2011, that politicians got direct access to their audience without the mediation of press [1]. Soon it was revealed that the influence of social media on politicians is no less than backward. Electoral campaigns are subjected to mediatization, namely, have to consider media logic and social media logic [2].

Therefore, politics became entertainment, and vice versa, their borders are blurring [3]. Funny pictures and videos played a significant role since the case of the Obama 2008 campaign [4]. This logic further developed after the Trump 2016 campaign. In his Twitter account, about half of the tweets were focused on the criticism of his opponent Hillary Clinton, and only other half – was dedicated to social issues and personal stories. In his campaign the text-only tweets about politics were more popular, then other types of content. Regarding Clinton, in her campaign were more efficient tweets with multimedia, emotional and personal [5].

Hannan claimed that 'If television turned politics into show business, then social media might be said to have turned it into a giant high school, replete with cool kids, losers and bullies.' [6]. He compared Obama with the school president, who spoke about his musical tastes, and Trump with a bully, who rules on the backyard.

Learning 2016 campaign and other examples of such logic, researchers found a lot of peculiarities of electoral campaign strategies in the SNS. This communication instrument facilitates agenda setting, fake news sharing and helps to create 'filter bubbles'. But only under the circumstance of a politician`s correct strategy [7].

## 1.2 Agenda setting in social networks

Agenda setting effect of Twitter accounts belonging to politicians and journalists was approved by scientists but to varying degrees. Rogstad (2016) found out that in America Twitter users more often catch up issues of the mainstream media than vice versa [8]. However, Twitter users play a significant role in critics of mainstream media. I other societies, for example in the Netherlands, during election campaigns interplay between mainstream and social media has been found commensurate [9]. Meanwhile, the strength of such influence depended on the politician`s position in the electoral roll, his age, gender and so on. Some investigations have shown a weak correlation between the agendas of different candidates [10]. So the interplay level may vary widely subject to specific circumstances.

As Facebook in the USA usually is not the part of mainstream political discourse, its influence on the people`s 'most important problems' is more important for the people who don't show much interest in the politics [11]. It is an example of the incidental news exposition. Opinion leaders have a major role to play in such conditions. Candidates` Instagram also has its own influence on the public agenda but usually with a very low level of coverage [12].

**Electoral issue ownership and its outcome**.

The party`s or candidate`s agenda is usually the mix of two approaches: reaction on the current actual problems of the public agenda, and issues ownership when a party tries to emphasize its skills in the specific problems solving [13].

The choice of the issue to own is very sophisticated. Sometimes it is possible to co-opt the competitor's agenda by offering extraordinary solvation of its problems [14] or by just reframing it [15], as Bill Clinton has done in 1996 with the crime problem owned by Republicans [16]. Therefore, in the issues of the American elections owned by two candidates usually, coincide for 30% to 50%.

Typically, a person's vote decision is proportionate to her subjective importance of some particular problem owned by some party [17]. Nevertheless, the 'punishment mechanism' might work, when it is widely assumed that the party failed in this problem-solving.

**Filter bubbles and elections.**

One more problem that should be considered in our research is the problem of the information variety. There are two ways for ordinary social media user: to take advantage of the wide diversity of information sources, or, instead, to select media and content that reinforce his existing beliefs, creating so-called filter bubbles or echo chambers. That is shown using the examples from American elections [18]. According to this, in Facebook echo chambers are stronger than in Twitter: supporters of Republicans usually got only not more than 5% of political content that does not match their beliefs, for Democrats supporters this number is 8%.

Researchers have ambiguous answers on the question about the echo chambers power. Thus, on the one hand, they proved, that even without any social or technological filters, echo chambers emerge as a consequence of cognitive mechanisms [19]. On the other hand, Dubois and Blank found out that the higher a person's level of political interest the less likely they are to be in an echo chamber [20]. Also, the filter bubble effect in Twitter moderates by the offline media consumption, i.e. TV watching [21].

There is even a study that shows: filter bubbles may prevent from additional polarization. When people are forced to read the Twitter accounts of politicians from the opposite camp, they tended to become more convinced in their prime political outlook, especially if this outlook was Republican [22].

But in any case, Echo chambers along with negative messages, fake news, and conspiracy theories became one of the recognizable features of the 2016 US presidential campaign [23].

**Social Networks in Ukraine and their influence on politics.**

Ukrainian social networks became a popular research object after the Revolution of Dignity beginning in 2013 and further, with the start of Ukrainian-Russian war, and Russia`s information intervention. These two interested researchers in fields of communications, political science, and sociology. It was found that Facebook became the working environment for Ukrainian protesters, and helped them to make new kind of protests, horizontal, without the core organizational structure [24]. Researchers agreed that SNS didn`t cause the protests, but helped to organize it by facilitating and acceleration of the information exchange and coordination of actions [25]. They also allowed to shape the identity discourse and increase the protester`s motivation [26]. SNS enlarged the spatial and social scope of the protests. There were discussions about the usefulness of online-only protest participation in Ukraine. Early researchers claimed that this type of activity led to protesters demobilization [27]. But then other opinion appeared. Surzhko-Harned and Zahuranec found that online participation helped to frame the revolution and to assure the protesters in their advantage [28].

In a few months, Russian intervention started along with the huge-scale information intervention. This gave to the scientists the opportunity to study propaganda and information warfare together with the self-defense practices developed by Ukrainian activists. The most interesting thing noted by scholars is the privatization of counter-propaganda and its shifting to the public sector [29]. This is a traditional function of the state, nevertheless, Ukrainian activists started crowdsourced online projects in fact-checking and propaganda revelation, international communications and creation of strategic narratives. Some of those activities were remarkably efficient, much more than state efforts [30].

All these circumstances led to the formation of a solid core of politically and socially active social network audience and increasing the connectivity level inside this core. Facebook, which is used in western countries mostly for entertainment, became a huge space for political communication. Information struggle made very influential dozens of so-called 'top-bloggers' who had different political positions with one common feature: patriotism [31]. The other parts of the worldview could vary, for example, a perspective on the economic and cultural issues. That is because Ukraine has Five different competing national narratives with different groundings [32]. This unit of opinion leaders was very efficient in the information self-defense during the Russian intervention. Fake news and Russian propaganda were neutralized by narratives spread by those influential users.

At that time Facebook was only third popular SNS (or fourth, considering YouTube as a social network) [33]. Leader positions hold Russian networks Vkontakte and Odnoklassniki. The first was used more by the younger audience for entertainment and to a less extent for political communication. The second was for an older audience and rarely belonged to politics. Both those networks were banned on May 15, 2017, as a threat to national security. Results of the blockage were contradictory. On the one hand, users quickly found out how to get around the block. Especially it is the case of Vkontakte users, who were younger and more technically advanced. On the other hand, the number of users and, especially, of content started to decrease steadily [34]. First of all, it`s about Odnoklassniki users most of who as older people didn`t succeed in the ban bypassing.

Most of the users who abandoned Russian SNS switched to Facebook (older people) and Instagram (younger generation) [34]. Alongside it, messenger mobile applications gained huge popularity in all age groups, especially Viber, used by 95% of smartphone owners, and Telegram [35]. They often substitute 'classical' SNS in spite of Viber don`t have an opportunity for public communication, so its users mainly use private groups for the communications. And the Telegram offers very poor options for feedback and interactions.

As a result, the most recent statistics of SNS use showed in March 2019, that 50% of adult Ukrainians use Facebook, 30% - YouTube, 27% use Instagram, 10% - Vkontakte, 7% - Twitter, and 6% - Odnoklassniki [36].

**Ukrainian 2019 presidential elections.**

Ukrainian 2019 presidential campaign was unusual. First of all, because nobody couldn`t predict for sure its outcome neither six months before its first round nor in a one month or even one week to it [37]. According to polls, five or six candidates had very similar levels of support and the share of people who didn`t make their choice ranged between 8% and 20% [38]. So, understanding of Ukrainian political environment seems to be crucial for the understanding of the research questions and results.

2019 elections also became outstanding by the recording number of candidates: there were 39 names in the ballot. Therefore, in the victory of three candidates, people believed more than in others [39]. They were: incumbent president Petro Poroshenko, former prime minister and political prisoner Yulia Tymoshenko and the comedian Volodymyr Zelenskyi. All three had very different agendas and very different stories to tell people.

Poroshenko told a story of national renovation. He armed himself with the patriotic agenda in spring of 2017 when he saw the critical reduction of his support. He started to patronize the laws about the Ukrainian language cultivation, about the historical memory of fight for Ukrainian independence. As a part of this, in April 2018 he launched the new project – the campaign of accession to independence of the Ukrainian Orthodox church. Until its successful ending in January 2019 this story was one of the most salient in the Ukrainian media and social networks, even for atheists and agnostics [40]. In parallel, he emphasized large-scale economic and social reforms he performed. Therefore, a lot of investigations revealed the facts of corruption in his inner circle, particularly in the defense sector, spoiled the impression of his motto 'The army, the language, the faith!'.

Yulia Tymoshenko told a story of the impoverishment of Ukrainian people. She promoted the problem of utility rates rising (so-called 'rates genocide'] in tandem with public exposure of incumbent`s corruption. Along with those negative messages, she promulgated the program of innovative economic growth.

Volodymyr Zelenskyi was not a politician. Before January 2019, nobody could say for sure whether he will participate in the elections or not. He was a well-known and successful comedian whose the most popular character was the president of Ukraine in the comedy series named 'Servant of the people'. The plot is that ordinary school teacher became the president due to the viral YouTube video and started to change the country and to rid it of oligarchs. After the start of the campaign, Zelenskyi engaged several liberal politicians to his team who negotiated with foreign partners of Ukraine. Therefore he avoided any specific statements about any political and economic issues, so he enjoyed the support of peoples with opposite expectations [41]. People also were worried about the ties between Zelenskyi and Ukrainian oligarch Ihor Kolomoiskyi whose channel broadcasted many of his shows.

Indeed, Zelenskyi, Poroshenko, and Tymoshenko became the three leaders of the first round with the outcome of 30,24%, 15,95%, and 13,4% respectively. So, the first two became the participators of the second round on April, 21$^{st}$. Anyone of highest-rated candidates didn`t express his support to those two winners because of the advent of parliamentary elections in the autumn of 2019.

Zelenskyi and Poroshenko started their new campaigns and tried to persuade people to make the second choice. We are going to look at this stage of the campaign, which led to the smashing defeat of Poroshenko. He earned the 24,45% compared to 73,22% of Zelenskyi. International monitors noted elections freedom and compliance with international standards.

So, we can see that election participators were succeeded in the agenda-setting of particular issues and in the issue of ownership. Two types of issues were ideological: Poroshenko 'owned' the patriotic agenda, Tymoshenko – the social one. All the incumbent`s critics made more salient the corruption issue which is not specific to any ideology. And the most interesting was the case of Zelenskyi. We may describe his campaign as an example of the non-agenda setting. This term was used for the situation when entertaining media by refusing to cover important issues give their readers the impression of the absence of any problems [42]. So, in the context of elections, we may use the concept of *non-agenda ownership*.

## 2      Hypotheses and research questions

As the social network services became so effective in the messages providing and the audience mobilization, they inevitably served as one of the battlefields of 2019 electoral campaign. Not only politicians paid huge attention to the social network agitation but also ordinary people discussed their choices and electoral news very actively. Therefore, the audience of different SNS separately and all the SNS cumulatively is not equal to all electorate and so may have a distinctive balance of two candidates` supporters. Moreover, the influence of different networks might be incomparable.

Despite the set of technically available methods, first of all, the content analysis, we could not determine the exact shares of two candidates` supporters. Therefore, we were able to compare their activity by the number of posts with candidates support, and popularity of these posts. So, the hypotheses are:

As these Ukrainian elections were the first after the blocking of the Russian SNS and dramatical changing of their usage patterns, and also the first after USA presidential elections known by their microtargeting practices, it was interesting to look at the patterns of the electoral of different users groups.

**RQ1.** What were the peculiarities of behavior of two candidates` supporters in the social networks?

**RQ2.** What were the peculiarities of behavior of two candidates` supporters in the electoral discussion in social networks?

Knowing the agenda of both candidates, we tried to compare electoral choice with users` worldview. It is not a simple issue to measure ideology in Ukrainian society. In the case of USA researchers usually use only one dimension: from left to right. That is not the case of Ukraine. It is possible for Ukrainians to be simultaneously socialists and nationalists, or nationalists and LGBT rights defenders. So we had to use three dimensions, based on the investigation of the Ukrainian national narratives: economic, foreign political and socio-political outlook. A detailed approach to the determination of these positions described below in the 'Method' chapter.

Thus, we know that 'liberal' issues aimed at markets freedom were slightly present in the campaigns of both candidates. So, we could assume two mutually exclusive hypotheses:

**H1a.** Public support of liberal ideas will positively predict the support of Zelenskyi rather than Poroshenko.

**H1b.** Public support of social ideas will positively predict the support of Poroshenko rather than Zelenskyi.

Defense, language and other patriotic issues were owned completely by Poroshenko, so we expect that:

**H2a.** Public support of patriotic ideas will positively predict the support of Poroshenko rather than Zelenskyi.

**H2b.** Public support of pro-Russian and cosmopolitan ideas will positively predict the support of Zelenskyi rather than Poroshenko.

No one of the two candidates didn`t pay significant attention to human rights issues. Therefore, taking into account traditional interconnection between the patriotic and conservative issues, we assume:

**H3a.** Public support of traditional outlook will positively predict the support of Poroshenko rather than Zelenskyi.

**H3b.** Public support of modern outlook will positively predict the support of Zelenskyi rather than Poroshenko.

In parallel, we tested the Ukrainian 2019 presidential campaign on the three main problems of the 2016 USA presidential campaign. It is the 'filter bubbles' phenomena in social networks, negative messages abuse and usage of the fake news and conspiracy theories.

**RQ3.** Was the filter bubbles phenomena essential in this campaign?

**RQ4.** Supporters of which candidates were more liked to be in the information bubbles?

**H4.** Negative messages were more salient in the discussion regardless of the side of this discussion.

**RQ5.** Where the fake news and conspiracy theories widely used in the campaign?

## 3  Method

For all those goals, we have used the content analysis for the data gathering and the discourse analysis for the data interpretation.

### 3.1 Analysis period.

April 11-15, 2019. On this time between the first and the second round of elections, the discussion was very active because two finalists diligently worked on the supporters` involvement. On April 11 Poroshenko unexpectedly came to the «Right for the power» TV show and had a heated phone exchange with Zelenskyi. Discussions continued about the format of the debate, about the provocative billboards with the incumbent president`s symbolic and about the candidates` drug test. On that time survey found that 20% of voters hadn`t yet made their choice.

### 3.2 Analysis sample.

To receive the widest range of web content, we used one of the most comprehensive commercial social media monitoring systems throughout the post-soviet space – YouScan. This system is usually used by PR-managers for the 'social media listening' cases and covers publically accessed content from all popular social networks to a significant degree. Therefore, it can be assumed, that the keyword search in this system able to find the significant share of all publically accessed content with those words.

We took posts, but not the comments or reposts in the six social networks often used by Ukrainians: Facebook, Instagram, Odnoklassniki, Twitter, VKontakte and Youtube. In total, during the period under study, we found 297,771 posts which contained the words «Зеленський», «Зе», «Порошенко» and «Порох», and had at least one interaction (like, «emotion», repost/retweet or comment).

Among this amount we randomly selected 1000 posts for the analysis. As a result of coding, we received 958 posts recognized as relevant, that are really devoted to candidates. Therefore, the error does not exceed 3%. We may consider that the share of posts from each social network in this sample is proportional to their shares in all electoral social media information flow.

Should be clarified, that, unlike western users, Ukrainians provide the largest share of social media political discussions in the public mood. We have no clear quantitative evidence of this fact; however. Indirect evidence may be considered, that overwhelming part of the Facebook shares of any political online publication is usually accessible in the Facebook search. This knowledge is widely used by political parties in their social media monitoring.

### 3.3 Coding.

Coding was conducted manually by the three trained coders and was held in three stages:

1. Coding of the content of posts in the sample. Thus, we identified the presence of the critics and support of both candidates in the posts and messages about the candidates which were delivered within these critics or support.

2. Coding the comments under these posts. Here we also determined the presence of critics or support of both candidates.
3. Coding of other content in the profiles of the authors of sample posts. We counted posts about elections among the last 30 posts in the public feed of these users. We checked the presence of personal content – life photo or stories, as well as posts showing the worldview of the people. We had three parameters of such the worldview: liberal/social, patriotic/cosmopolitan (or pro-Russian), modern/conservative outlook.

User outlook was judged liberal in a case he supported the tax cut or the tax system reform, enterprise freedom, deregulation, and the market-based utility rates. Social was considered to be the position of users who supported rate decrease, wages and pensions boost, etc. The pro-Russian outlook was not separated from the cosmopolitan because of the national circumstances and was determined by the support of Russia`s actions in Ukraine, dissatisfaction with the language bill considered in the Ukrainian Parliament on that time, by the critics of Ukrainian army and so on. And vice versa, the patriotic outlook of the user was determined by the support of the language bill, Ukrainian orthodox church independence, Ukrainian soldiers, etc. conservative or progressive outlook was identified by the attitude to LGBT, women and national minorities rights, family values and so on.

In addition, the monitoring system automatically determined the number of post`s interactions, the user`s gender and region where he lives, as indicated by himself.

In most figures, we defined the support of a candidate in the post as the presence of positive message about him or negative message about his competitor. Distinguish of these two types of messages is taken to account only in the Fig.7.

## 4   Results

Undoubtedly, the activity of both sides of the discussion was not proportionate to neither election outcome (Fig.1) nor the relative popularity of the social networks (Fig.2). There was almost equal (50%:49%) number of posts from supporters of two candidates. Moreover, posts of Poroshenko`s supporters have got 80% of all interactions (comments, likes, 'emotions', shares and retweets). This data differs significantly from the outcome of the election (24,45% for the incumbent). As we will saw below, this superiority was achieved by only several posts of well-known users.

We cannot determine what has affected more for the Poroshenko`s result: the greater activity of Poroshenko`s supporters or their greater share in social networks in comparison to all Ukraine population. That could be determined by the polls, and our aim is to describe discussion rather than every its participant.

The choice did not depend on the gender: we saw almost equal shares of the candidate`s support from men and women. Simultaneous critics or support of both candidates were found only in the less than 1% of posts. So, most of the posts dedicated to elections were written by users who had already done their choice.

In the frame of different social networks, results are even more interesting. The distinction between common and electoral usage of different social networks may be explained by the traditions of social media consumption in Ukraine. Despite the 27% of Instagram penetration, electoral posts in this SNS occupy only 0,52% in all sample of political social media content. Unlike it, Facebook has 92% of all electoral posts unlike to its 50% penetration in Ukrainian society. Other SNS were also participated less in political discussions than in everyday people`s life.

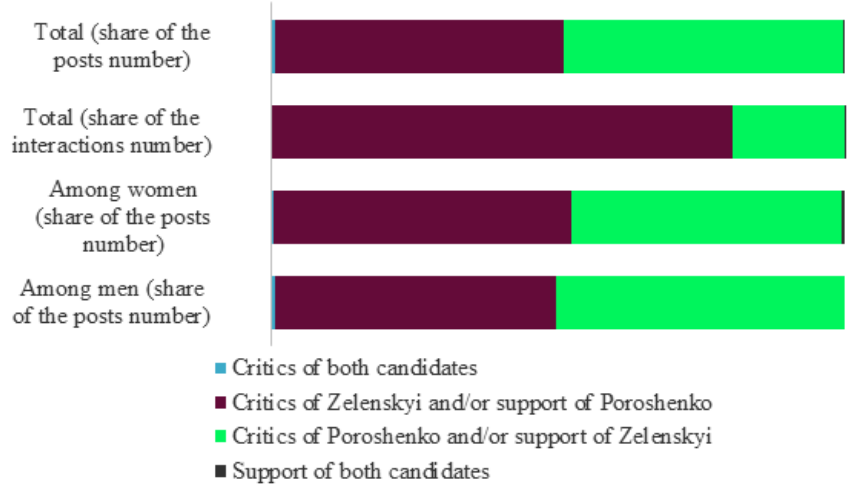

**Fig. 1.** Total and gender distribution of content supporting or criticizing candidates; number and share of posts.

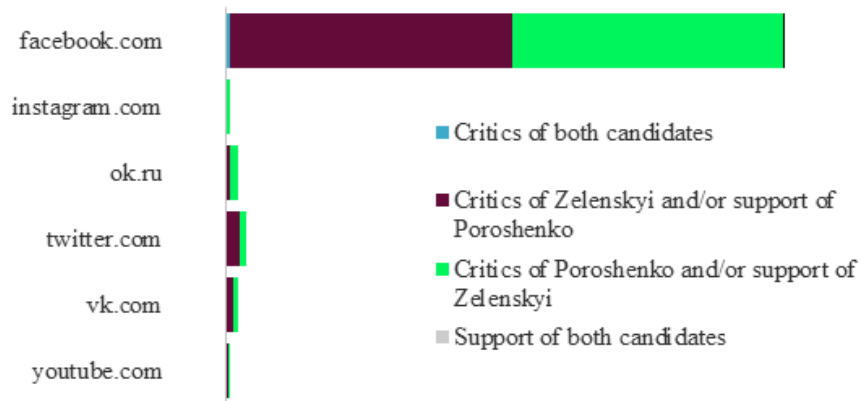

**Fig. 2.** Number and share of posts of candidates` supporters in different social networks; number and share of posts.

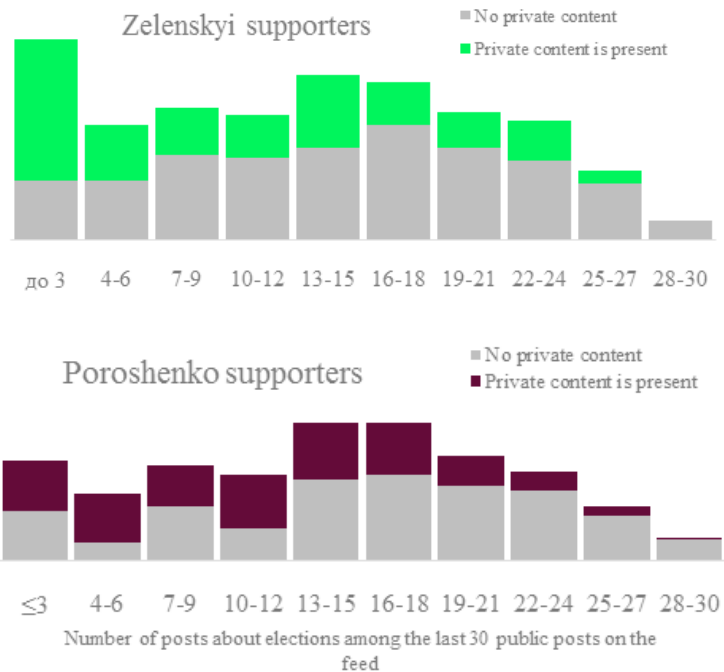

**Fig. 3.** The presence of the personal content in the profiles and posts of two candidates` supporters depending on their political expression activity; a number of posts by people (not communities).

Our sample doesn`t contain any Instagram posts with support of Poroshenko or critics of Zelenskyi. Its activity in the political discussions was even less than in banned Russian SNS. The first of them, Odnoklassniki, was represented in the sample mostly by the posts from the occupied territories of Ukraine.

The support of Zelenskyi in Vkontakte was also found mostly in the pro-Russian groups, therefore, there were some posts in this SNS from ordinary users who supported Poroshenko. Twitter content contained only the posts from the media accounts, not from personal blogs. All YouTube videos in the sample were posted by popular channels.

Definitely, it does not mean that the communication in this SNS was the most effective. Content in other services could be more persuasive, for example, due to the filter bubbles formation or for a greater number of views of each YouTube video.

Regarding the behavior of users, we paid attention to the two parameters: the number of posts dedicated to elections in the last 30 public posts (the level of politicization) and the presence of the private content (live photos or stories) in the public part of Facebook. This second figure may help us to determine the prevalent type of User`s experience. Publicity in the SNS is usual to 'old' Facebook users who joined this service before 2014, are active in it and have more or less enough knowledge about political life, as well as for users who use SNS first of all for the entertainment instead of political activity. This kind of users particularly used actively Vkontakte and only after its ban became more active on Facebook. And vice versa, absence of personal content might be explained by rare, inactive and timid SNS usage or traditions of the mostly private SNS usage of the former Odnoklassniki audience.

So, nearly 60% of both Poroshenko and Zelenskyi supporters had no publicly accessible private information. Also, there were two most widespread types of users: who wrote only 3 or less electoral posts among the last 30, and who dedicated to elections approximately half of all posts. The first type was a bit more characteristic for Zelenskyi`s supporters and the second was for Poroshenko`s ones. But the most interesting is a comparison to these two parameters.

In the category of weakly politized users (**3 or less electoral posts**) Zelenskyi`s supporters were usually people with a lot of personal and entertaining content, unlike the Poroshenko`s supporters who were often looked like inactive users. People with **7-9 electoral posts** were very active it both camps and had a substantive amount of friends. Therefore, Poroshenko`s supporters in this group paid more attention to social issues than to entertainment. The politically active users with **10-18 posts dedicated to elections** were very different for two candidates. This group of Zelenskyi`s supporters mainly had profiles full only by reposts. And the majority of Poroshenko`s supporters were active users with a lot of personal content. Authors of **19-21 posts about elections** regardless of their political position usually have only reposts in their profiles, therefore Zelenskyi`s supporters had more personal content, and Poroshenko`s supporters in average had a greater number of friends and more detailed bios.

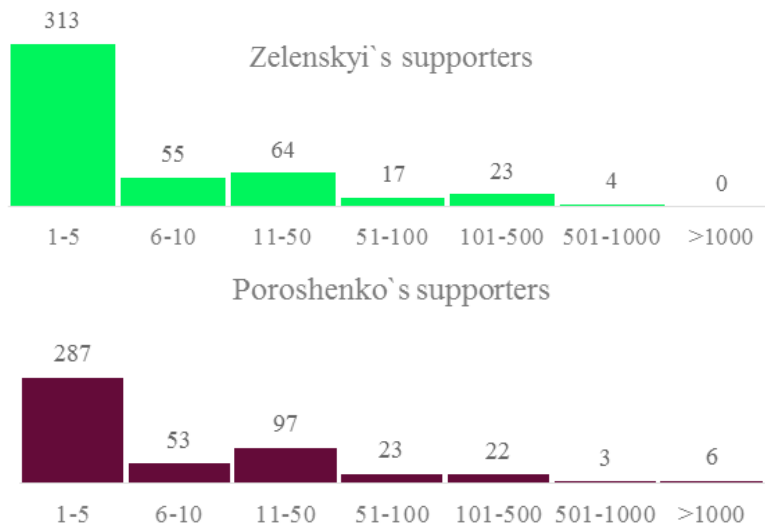

**Fig. 4.** Frequency distribution of interaction numbers of the posts of both candidates` supporters; a number of posts by people (not communities).

So, we can see that Zelenskyi had two 'cores' of the Social networks electorate. The first consisted of active users of Facebook, Instagram, and Vkontakte who are not much interested in politics and usually use SNS only for joy. And the second 'core' was formed by people who are not indifferent to politics but usually don`t participate in mainstream political discussions. The 'core' of Poroshenko`s active supporters were people, highly engaged in political issues, produce a lot of personal content and use to communicate in SNS. Mostly they are users with great experience in Facebook life.

Histograms of the interaction numbers of the posts (Fig.4.) look almost equal. The only important difference is in the last bin, posts with more than 1000 interactions. Zelenskyi`s supporters had not so popular posts as so-called 'top-bloggers' who supported Poroshenko. And it was only this bin which provided a Poroshenko`s supporters leadership by the number of interactions.

Ideological issues had a very different salience in the electoral period of time. Social rhetoric was boosted by Tymoshenko and her less successful competitors. Patriotic issues – by Poroshenko. So these two types of indicators most often happened in the sample of profiles. And vice versa, public demand on the liberal reforms was high right after the Revolution of dignity but then became less prominent. Pro-Russian outlook became indecent after the beginning of the war with Russia. The controversy between the bearers of traditional and modern values was very divisive for the Ukrainian society in the 2017 year, but in the advent of elections faded away.

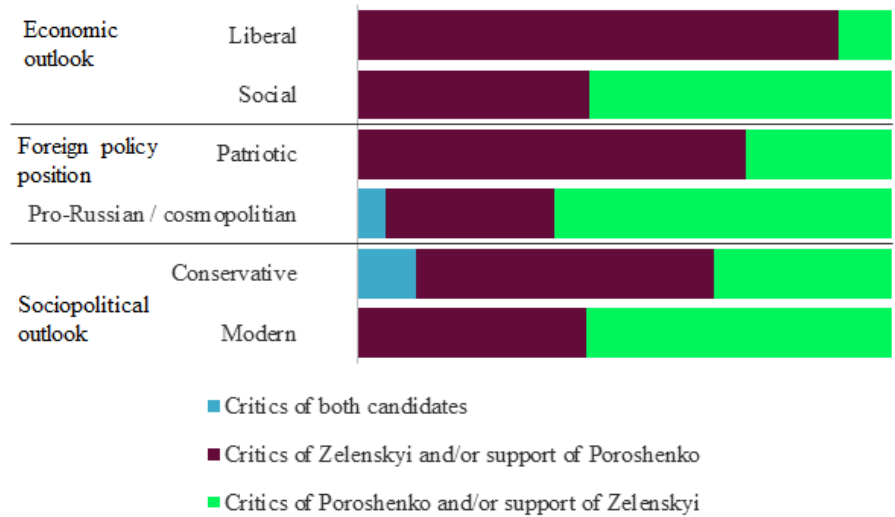

**Fig. 5.** An electoral choice of social media users depending on their world view; number and share of posts by people (not communities).

So, as seen in the Fig.5, Our observations supported the **H1b** and didn`t support the **H1a** position. Moreover, we found the equal number of the social issues defenders among both candidates` supporters.

**H2a** and **H2b** were confirmed in general. Therefore near the 25% of 'patriots' supported Zelenskyi contrary to the Poroshenko`s issue ownership. Furthermore, we found even the bearers of cosmopolitical worldview and defenders of bilingual socie-ty among the Poroshenko`s supporters. In other words, they supported the incumbent in spite of his language policy.

**H3a** and **H3b** were weakly confirmed: there was a slight difference in the choices of users with a modern and traditional outlook.

Critics of both candidates simultaneously were detected from some pro-Russian and conservative users.

For the **RQ3.** and **RQ4.** we counted the number of posts with at least one comment and looked on these comments orientation (Fig.6.). Almost half of the Zelenskyi`s supporter`s posts had no comments of the opponents. For Poroshenko`s supporters, this share was the one third. That is not the accurate share of the people in the bubble (see 'Discussion') but gives us a sense of the scale. The most homogeneous SMS was Instagram: there weren`t any Poroshenko`s supporters posts as well as their comments.

Users worldview was also one of the factors influenced on the bubble`s appearance. All posts from 'liberals' regardless of their electoral choice had the comments of opponents. In other words, 'liberals' inclined to have more diverse and pluralistic environment. Therefore, attention to social issues does not influent on the 'bubble'.

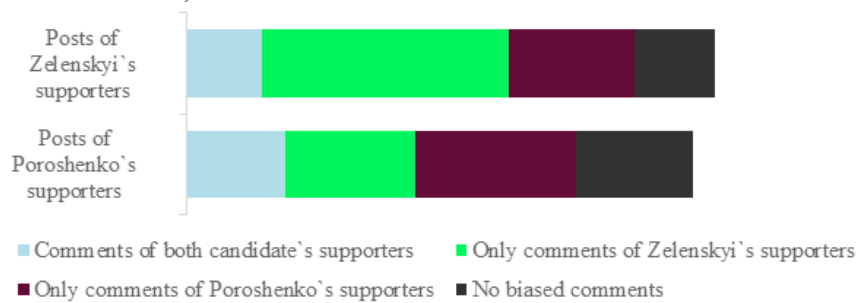

**Fig. 6.** Adherence of the comments toward candidates under the posts of supporters of those candidates; number and share of posts by people (not communities).

The opposite situation is in the dimension of the socio-political outlook. Both conservative and modern worldviews didn`t support the discussion. Electoral posts from LGBT rights supporters had no comments from other candidate`s supporters as well as posts from traditional family defenders.

In accordance with **H4**, negative messages indeed were more widespread in both campaigns (Fig.6.). In other words, both users described their candidates as 'less evil' rather than as 'good boy'. Therefore, the picture is changing with the increasing of the number of interactions. The half of the most prominent posts (over 500 interactions for Zelenskyi`s supporters and over 1000 for Poroshenko`s ones) contain the positive messages. With the decreasing of the popularity of the post, the share of positive ports also gradually decreases to 21% among posts with three or fewer interactions.

It is noteworthy that the share of negative and positive messages does not depend on the account politicization, i.e. the share of electoral posts among the last ones.

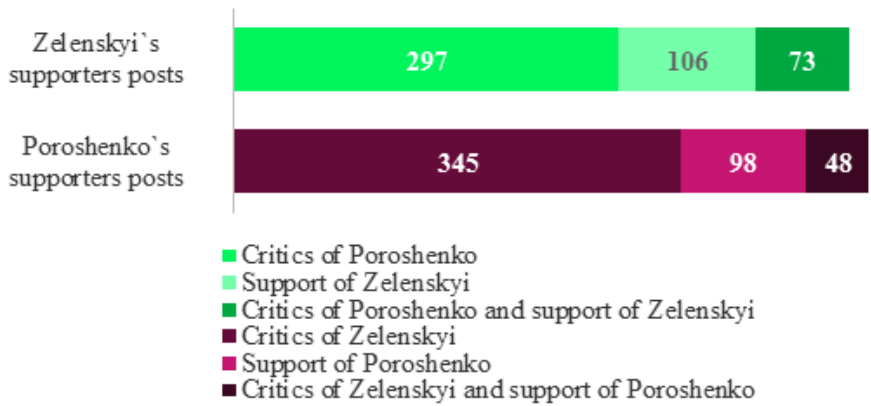

**Fig. 7.** Share of positive and negative messages in the posts of both candidates` supporters; a number of posts.

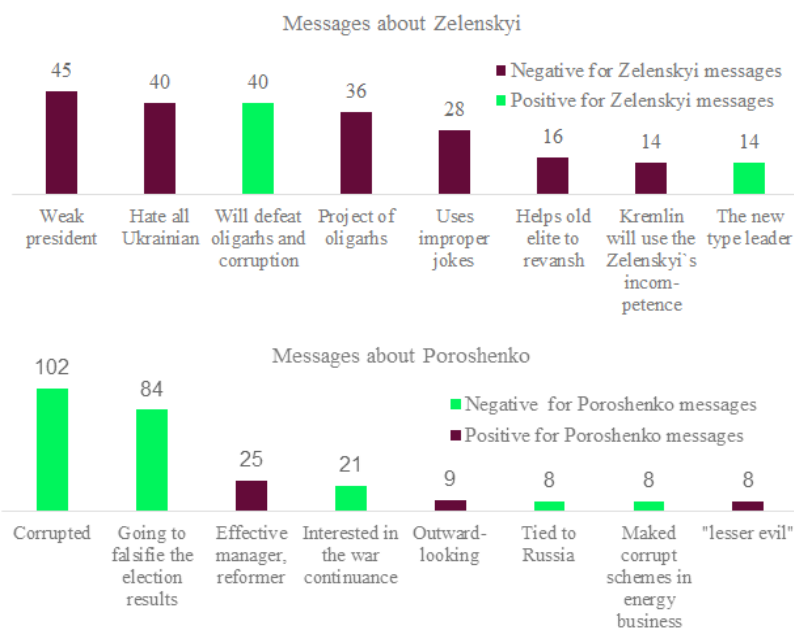

**Fig. 8.** The most widespread in the social networks messages about candidates, number of posts.

Share of positive varies also by the regions of Ukraine. It may be explained by different people`s interests in different regions along with the intentional targeting of different messages to different audiences. A good example is a Western region where the share of posts with positive messages about Zelenskyi was 49% of all messages of his supporters – much more than on average for the country.

The most popular messages about two candidates we give in the Fig.7. We have determined that messages of Poroshenko`s supporters were facts of value judgments. The situation with Zelenskyi`s supporters' messages is pretty different. At least one positive message and two negative ones were recognized as fakes by leading Ukrainian fact-checking information campaign "Behind the news". For example, further denied statements about the future appointments of known persons on the state senior positions. Or conspiracy theory about deliberate defeats for Ukrainian army aimed at higher coal prices.

All those fake messages were boosted in social networks via a plethora of news websites without a certain audience or reputation.

## 5 Discussion and conclusion

### 5.1 Agenda setting in the highly diverse ideology system

Engagement of social networks users into electoral campaigns became very salient. No doubts, different countries and communities have their own typical patterns of SNS use. Ethnographers noticed that different cultures have very different traditions of social media behavior [43]. That`s why in Ukraine, which has a very active 'core' of social media users, divided by three different ideological dimensions and used to post political content publically, this engagement became prominent. We see that 64% of people who wrote at least one public post about elections dedicated to this topic more than one-third of all their posts in this period of time. They tried to persuade their friends and influent their choice. In other words, the popular metaphor that elections are like horse-racing now modified: the horse racing became horse supporting.

**Non-agenda ownership and its interplay with ideological outlook.**

We see clear evidence of the bigger salience of issues owned by two of the three leading candidates: Poroshenko and Tymoshenko. Users wrote much more about 'patriotic' issues, and, particularly, about 'social', then others. So, we may assume, that writing about issues like those two is not a testimony of user`s real worldview. Those issues may be just imposed and temporary seems important to people. We think so because the set of 'socialists' has the same indicators as for the whole sample: the same share of their electoral choice, the share of negative messages and so on. So, this sample seems almost random.

Nevertheless, ownership of patriotic issues does not lead to automatic voting of people who share patriotic views. Only 73% of users who had posts with patriotic position actually supported Poroshenko. A quarter of them has made the opposite choice as they believe that Poroshenko failed to solve national problems. These people would not vote for the Poroshenko`s competitor if he had been exactly pro-Russian. But the non-agenda setting made Zelenskyi the 'neutral' choice. It is an example of a 'punishment mechanism' described by Bélanger and Meguid [17].

But Zelenskyi`s supporters were so diverse that we may find the significant amount of them in any ideological group of all three dimensions. Suppression of his ideological platform could not be just a populism, because populism usually has its ideological disposition: left, right or, at least, centric. Zelenskyi avoided such positioning and had no competitors in the non-agenda ownership.

Further, we may assume that other ideological issues appeared in the SNS and were not boosted seem to be indicators of real worldview and users` concerns. Noteworthy among them is the liberal position. Despite the attempts of participants of Zelenskyi`s team to announce the liberal reforms, his electorate didn`t care about the tax reform or the land market. 90% of the 'liberals' called to vote for Poroshenko.

In addition, we noticed that users who wrote about liberal issues were not in bubbles at all. They always had the comments of their opponents under their electoral posts. Probably, the liberal outlook is much more open to communication with the people with an alternative view. Quite the opposite situation is with people who took care of both modern and conservative issues. They all had no comments from another candidate`s supporters. This corresponds with Halberstam and Knight result as they found out that Democrats are less likely to be in the bubbles than the Republicans.

### 5.2   Similarity with Trump`s 2016 campaign

In addition, we noticed the significant similarity between the problems typical for two campaigns: American and Ukrainian. That is a landmark conclusion for the Ukrainian communication activists who were proud of the ability of Ukrainian social network community for the fake combating. This ability was valid for the period when Facebook users mainly were active people connected to each other. And now all manipulative methods developed particularly by Russian trolls ant tested on Trump`s elections are suitable for Ukraine. So, this country became less specific.

**Social network core as a huge filter bubble.**

SNS seems to be hybrid systems with non-linear electoral output and constantly evolving patterns and rules. There is an ambiguous influence of politicians, their agendas and digital strategies on the users.

The SNS strategy of Ukrainian incumbent president Petro Poroshenko was developed by several years. Its main goal was the deliberative or manipulative control of opinion leaders and their first-, second and third-level agenda.

As the core of Ukrainian Facebook community was formed as an aftermath of the Revolution of Dignity, it had a significant share of politically and socially active people who are able to be simply mobilized to political actions – i.e. revolutions or voting. The main genre of opinion leaders communication was a comparatively long emotional text. The main assessment of their performance was the number of interactions reached by their posts. That is because interactions characterized mobilization much better than passive contacts with the content usually named 'views'. This strategy helped Poroshenko`s team to reach 80% of all the interactions in the electoral information flow in SNS. Some posts had dozens of thousands likes, shares and comments.

But it was not the case of Zelenskyi. He chose another strategy and reached the audience poorly linked to the `core` – so-called social networks 'periphery'. Those people never used to read the opinion leaders` posts. Some of them joined Facebook after experience in Russian SNS. They preferred pictures rather than long texts. But they have written nearly the same number of electoral posts as the 'core'. Yes, the number of interactions of posts written by Zelenskyi`s supporters did not exceed 1000 in the sample. But they have spread a lot of information from Zelenskyi`s official YouTube channel and from a plethora of fake websites.

Regional Facebook groups of Zelenskyi`s supporters, as well as the groups in the text messengers, were widely used in this content dissemination. These groups also helped to target the content to the different segments of their audience.

By the way, Facebook announced the changes in its communicative politics and growing the importance of groups for more confident and protected communication. Therefore, our research testifies that group conversations remained nonetheless vulnerable to manipulations than the traditional open and direct pattern.

**The first similarity: filter bubbles.**

We have not measured accurately the share of people in filter bubbles because it requires the use of the social network analysis. Therefore, even a rough estimate shows that the prevalence of bubbles was high.

Our measurements showed that half of Zelenskyi`s supporters` posts had no opponents` comments. The same thing happened to the one-third of posts of Poroshenko`s supporters. But these are exactly the shares of posts, not of the users. We see that the majority of users wrote more than one electoral post. So, if disagreed commentators didn`t appear under the first their post, they may come under the second or third. Or just write their own posts which will be seen in the feed. This factor makes the bubble share smaller than our measurement.

On the other hand, we may only guess how many opponents have in their friend networks users whose posts didn't reach even one comment. The share of uncommented posts was two thirds. It sounds logical that if they had a significant amount of friends-opponents, either of them would probably argue. So it is likely that the absence of comments indicates that the authors of those unpopular posts were surrounded by rather like-minded friends. This factor makes the bubble share larger than our measurement.

**The first similarity: negative messages superiority.**

Negative messages were used in both campaigns between the two rounds more often than positive, especially with ordinary users posts. Nevertheless, posts with positive messages became popular much more often. We see two different explanations for this trend. The first is that this type of posts was written above all by the most popular users. The second is that positive content is more attractive to users. To determine which factor is more significant, we have calculated for each candidate`s supporters the ratio of the average number of their posts interactions to the average number of their followers. This ratio is 0,05696 for Poroshenko and 0,0061 for Zelenskyi. But taking into account only posts with positive messages enlarge these indicators to 0,1131 and 0,0097 respectively. So, positive content was after all more attractive.

**The third similarity: fake news and conspiracy.**

We didn`t dig deeply to unmask all the strategy of fake news usage by both candidates. But it was surprising for us that even in our short sample the fake messages from Zelenskyi`s supporters were in the lists of his TOP messages reached dozens of shares. That was the best evidence of the prominent role of fake news along with the non-agenda ownership.

## Acknowledgments

We are very appreciative to Fulbright program of the Institute of International Education and to the Center for the Content Analysis for the support in this research.

## References

1. Cooper M. Structured viral communications : the political economy and social organization of digital disintermediation. J Telecommun High Technol Law. 9(1), 15–80 (2011).
2. Skogerbø E, Karlsen R. Mediatisation and regional campaigning in a party centred-system: how and why parliamentary candidates seek visibility. Javnost. 21(2), 75–92 (2014). http://search.ebscohost.com/login.aspx?direct=true&db=cms&AN=97122729&site=ehost-live
3. Wheeler M. The mediatization of celebrity politics through the social media. Int J Digit Telev. 5(3), 221–35 (2014). http://openurl.ingenta.com/content/xref?genre=article&issn=2040-4182&volume=5&issue=3&spage=221
4. Hussain MM. Journalism's digital disconnect: The growth of campaign content and entertainment gatekeepers in viral political information. Journalism. 2012;13(8):1024–40.
5. Lee J, Xu W. The more attacks, the more retweets: Trump's and Clinton's agenda setting on Twitter. Public Relat Rev. 44(2), 201–13 (2018). Doi: 10.1016/j.pubrev.2017.10.002
6. Hannan J. Trolling ourselves to death? Social media and post-truth politics. Eur J Commun. 33(2), 214–26 (2018).
7. Conway-Silva BA, Filer CR, Kenski K, Tsetsi E. Reassessing Twitter's Agenda-Building Power: An Analysis of Intermedia Agenda-Setting Effects During the 2016 Presidential Primary Season. Soc Sci Comput Rev. 36(4), 469–83 (2018).
8. Rogstad I. Is Twitter just rehashing? Intermedia agenda setting between Twitter and mainstream media. J Inf Technol Polit. 13(2), 142–58 (2016).
   Kruikemeier S, Gattermann K, Vliegenthart R. Understanding the dynamics of politicians' visibility in traditional and social media. Inf Soc. 34(4), 215–28 (2018). doi: 10.1080/01972243.2018.1463334
   de Los Angeles Flores M. Inter-candidate Facebook Agenda-Building Effect of U . S . -Mexico-Border Web County Judgeship Election Campaigns. Norteamérica. 12(2), 1–25 (2017).

9. Feezell JT. Agenda Setting through Social Media: The Importance of Incidental News Exposure and Social Filtering in the Digital Era. Polit Res Q. 71(2), 482–94 (2018).
10. Towner TL, Muñoz CL. Picture Perfect? The Role of Instagram in Issue Agenda Setting During the 2016 Presidential Primary Campaign. Soc Sci Comput Rev. 36(4), 484–99 (2018).
11. Sides J. The origins of campaign agendas. Br J Polit Sci. 36(3), 407–36 (2006).
12. Sides J. The consequences of campaign agendas. Am Polit Res. 35(4), 465–88 (2007).
13. Petrocik JR, Benoit WL, Hansen GJ. Issue Ownership and Presidential Campaigning, 1952-2000. Polit Sci Q. 118(4), 599–626 (2003).
14. Holian DB. He's Stealing My Issues! Clinton's Crime Rhetoric and the Dynamics of Issue Ownership. Polit Behav. 26(2), 95–124 (2004).
15. Bélanger É, Meguid BM. Issue salience, issue ownership, and issue-based vote choice. Elect Stud. 27(3), 477–91 (2008).
16. Halberstam Y, Knight B. Homophily, group size, and the diffusion of political information in social networks: Evidence from Twitter. J Public Econ. 143:73–88 (2016). DOI: 10.1016/j.jpubeco.2016.08.011
17. Geschke D, Lorenz J, Holtz P. The triple-filter bubble: Using agent-based modelling to test a meta-theoretical framework for the emergence of filter bubbles and echo chambers. Br J Soc Psychol. 58(1), 129–49 (2019).
18. Dubois E, Blank G. The echo chamber is overstated: the moderating effect of political interest and diverse media. Inf Commun Soc. 21(5), 729–45 (2018).
19. Eady G, Nagler J, Guess A, Zilinsky J, Tucker JA. How Many People Live in Political Bubbles on Social Media? Evidence From Linked Survey and Twitter Data. SAGE Open. 9 (1), 1–42 (2019). https://journals.sagepub.com/doi/pdf/10.1177/2158244019832705
20. Bail CA, Argyle LP, Brown TW, Bumpus JP, Chen H, Hunzaker MBF, et al. Exposure to opposing views on social media can increase political polarization. Proc Natl Acad Sci. 115(37), 9216–21 (2018). DOI: 10.1073/pnas.1804840115
21. Pennycook G, Rand DG. Lazy, not biased: Susceptibility to partisan fake news is better explained by lack of reasoning than by motivated reasoning. Cognition, 1–12 (2018). DOI: 10.1016/j.cognition.2018.06.011
22. Bohdanova T. Unexpected revolution: the role of social media in Ukraine's Euromaidan uprising. Eur View. 13(2), 347–347 (2014). http://link.springer.com/10.1007/s12290-014-0314-6
23. Jost JT, Barberá P, Bonneau R, Langer M, Metzger M, Nagler J, et al. How Social Media Facilitates Political Protest: Information, Motivation, and Social Networks. Polit Psychol. 39(3), 85–118 (2018).
24. Onuch O, Pal MK. EuroMaidan protests in Ukraine: Social media versus social networks. Probl Post-Communism. 62(4), 217–35 (2015). DOI: 10.1080/10758216.2015.1037676
25. Onuch O. "Facebook Helped Me Do It": Understanding the EuroMaidan Protester "Tool-Kit." Stud Ethn Natl. 15(1), 170–84 (2015). DOI: 10.1111/sena.12129/abstract
26. Surzhko-Harned L, Zahuranec AJ. Framing the revolution: the role of social media in Ukraine's Euromaidan movement. Natl Pap. 45(5), 758–79 (2017).
27. Bolin G, Jordan P, Ståhlberg P. Information Warfare The Management of Information in. In: Pantti M, editor. Media and the Ukraine Crisis: Hybrid Media Practices and Narratives of Conflict. Peter Lang p. 3–18 (2016).
28. Sienkiewicz M. Open source warfare: the role of user-generated content in the Ukrainian Conflict media strategy. In: Media and the Ukraine Crisis: Hybrid Media Practices and Narratives of Conflict, p. 19–70 (2016).

29. ТОП-50 лідерів думок у соцмережах | Новое Время. (2016). https://magazine.nv.ua/ukr/journal/2597-journal-no-43/geroi-facebook.html
30. Korostelina KV. Constructing the narratives of identity and power: Self-imagination in a young Ukrainian nation. Lanham: Lexington Books. 260 p (2014).
31. Negreyeva I. USAID U-Media annual media consumption survey (UKR). (2016). https://www.slideshare.net/umedia/usaid-umedia-annual-media-consumption-survey
32. PlusOne. Instagram в Україні, (2018). http://plusone.com.ua/insta/
33. AIN.ua. Топ-5 мессенджеров в Украине: больше 90% пользуются Viber | AIN.UA ., (2019). https://ain.ua/2018/04/10/top-5-messendzherov-v-ukraine/
34. Research & Branding Group. Практика пользования соцсетями в Украине. , (2019). http://rb.com.ua/blog/praktika-polzovanija-socsetjami-v-ukraine/?fbclid=IwAR3LAJVpOwYPgDWdQ7uqvkWXrUnX8S_rkXry9nhmtuxsNUsinpNnn96zIvs
35. Наталія Судакова. Президентські вибори 2019. Що змінять голоси невизначених?, (2019). https://www.pravda.com.ua/articles/2019/03/21/7209775/
36. Public Opinion Survey of Residents of Ukraine. [cited 2019 May 13]. Available from: https://www.iri.org/sites/default/files/2019.1.30_ukraine_poll.pdf?fbclid=IwAR2G2xlN6YrBwxMftBXkBHEMmBDjbajvi7RN6bDbbaqAoo4J9ksy5tCLAWE
37. SOCIS. «Україна Напередодні Президентських Виборів 2019», (2019). http://socis.kiev.ua/wp-content/uploads/2019/01/Press_reliz_3_company.pdf
38. Zakharchenko A. Історія про томос: як ставилася найпопулярніша політична драма року, (2019). https://www.pravda.com.ua/columns/2019/04/1/7210952/
39. Центр Разумкова. 5 політичних дилем: що пропонують кандидати в президенти та обирають їх прихильники, (2019). http://razumkov.org.ua/images/Material_Conference/2019_03_19/2019_Prezent_ukrinform.pdf
40. Pingree RJ, Stoycheff E, Sui M, Peifer JT. Setting a Non-Agenda: Effects of a Perceived Lack of Problems in Recent News or Twitter. Mass Commun Soc . 2018;21(5):555–84. Available from: https://doi.org/10.1080/15205436.2018.1451543
41. Costa E. Affordances-in-practice: An ethnographic critique of social media logic and context collapse. New Media Soc., 20:3641 –3656 (2018).